\documentclass[12pt,preprint]{aastex}

\usepackage{color}
\usepackage{ulem}
\usepackage{amsmath}
\usepackage{hyperref}
\def\AB{\textcolor{black}}

\def\simlt{\lower.5ex\hbox{$\; \buildrel < \over \sim \;$}}
\def\simgt{\lower.5ex\hbox{$\; \buildrel > \over \sim \;$}}

\def\beq{\begin{equation}}
\def\eeq{\end{equation}}
\def\ba{\begin{eqnarray}}
\def\ea{\end{eqnarray}}

\def\bB{{\,\mathbf B}}
\def\bE{{\,\mathbf E}}

\def\bJ{{\,\mathbf J}}
\def\bv{{\,\mathbf v}}

\def\rhoGJ{\rho_{\rm GJ}}
\def\IGJ{I_{\rm GJ}}
\def\gthr{\gamma_{\rm thr}}
\def\RLC{R_{\rm LC}}
\def\Rout{R_{\rm out}}

\def\br{{\mathbf r}}
\def\Phipc{\Phi_0}

\def\gamLC{\gamma_0}

\def\bvrot{\bv_{\rm rot}}
\def\bOm{\boldsymbol{\Omega}}
\def\bmu{\boldsymbol{\mu}}
\def\Lsd{L_{\rm sd}}
\def\lav{\bar{l}}
\def\ldisp{\Delta l}

\def\Rs{R_{\star}}
\def\rL{r_{\rm L}}
\def\dNi{\dot{N}_i}
\newbox\grsign \setbox\grsign=\hbox{$>$} \newdimen\grdimen \grdimen=\ht\grsign
\newbox\simlessbox \newbox\simgreatbox \newbox\simpropbox
\setbox\simgreatbox=\hbox{\raise.5ex\hbox{$>$}\llap
     {\lower.5ex\hbox{$\sim$}}}\ht1=\grdimen\dp1=0pt
\setbox\simlessbox=\hbox{\raise.5ex\hbox{$<$}\llap
     {\lower.5ex\hbox{$\sim$}}}\ht2=\grdimen\dp2=0pt
\setbox\simpropbox=\hbox{\raise.5ex\hbox{$\propto$}\llap
     {\lower.5ex\hbox{$\sim$}}}\ht2=\grdimen\dp2=0pt
\def\simgt{\mathrel{\copy\simgreatbox}}
\def\simlt{\mathrel{\copy\simlessbox}}

\begin{document}

\title{Electrodynamics of axisymmetric pulsar magnetosphere with 
electron-positron discharge: a numerical experiment}

\author{Alexander Y. Chen, Andrei M. Beloborodov}
\affil{Physics Department and Columbia Astrophysics Laboratory,
Columbia University, 538  West 120th Street New York, NY 10027
%; amb@phys.columbia.edu
}

\begin{abstract}
We present the first self-consistent global simulations of pulsar magnetospheres 
with operating $e^\pm$ discharge. We focus on the simple configuration of 
an aligned or anti-aligned rotator. The star is spun up 
from zero (vacuum) state to a high angular velocity, and we follow the coupled
evolution of its external electromagnetic field and plasma particles using the 
``particle-in-cell'' method. A plasma magnetosphere begins to form 
through the extraction of particles from the star; these particles are accelerated 
by the rotation-induced electric field, producing curvature radiation and 
igniting $e^\pm$ discharge. We follow the system evolution for several
revolution periods, longer than required to reach a quasi-steady state.
Our numerical experiment puts to test previous ideas for the plasma flow and gaps 
in the pulsar magnetosphere.

We first consider rotators capable of producing pairs
out to the light cylinder through photon-photon collisions.  We find
that their magnetospheres are similar to the previously obtained
force-free solutions with a Y-shaped current sheet. 
The magnetosphere continually ejects $e^\pm$ pairs and ions.
Pair creation is sustained by a strong electric field
along the current sheet.  We observe powerful curvature and
synchrotron emission from the current sheet, consistent with {\it
  Fermi} observations of gamma-ray pulsars.

We then study pulsars that can only create pairs in the strong-field region near the 
neutron star, well inside the light cylinder.
We find that both aligned and anti-aligned rotators relax to the ``dead'' state with 
suppressed pair creation and electric currents, regardless of the discharge voltage.

\end{abstract}


%###################################################################

\section{Introduction}

The standard picture of pulsar magnetosphere assumes that it is
filled with plasma and  corotates with the neutron star with angular velocity $\Omega$
% (Goldreich \& Julian 1969, hereafter GJ). 
\citep[hereafter GJ]{goldreich_pulsar_1969}.  GJ considered the
aligned rotator (magnetic dipole moment $\bmu$ parallel to $\bOm$);
then it was generalized to inclined rotators.  The
plasma sustains the ``corotational'' electric field $\bE\approx
-\bv_{\rm rot}\times \bB/c$
(with $\bvrot=\bOm\times\br$), which implies the local charge density 
$4\pi\rhoGJ=\nabla\cdot \bE \approx -2\bOm\cdot\bB/c$.
A key feature of the GJ model is the electric current $\IGJ$ flowing out of
and into the star along the open magnetic field lines that extend to
the light cylinder $\RLC=c/\Omega$. GJ showed that the open field
lines are twisted and exert a spindown torque on the rotator.  The
circulating current is $\IGJ\approx \mu \Omega^2/c$, and the
corresponding spindown power is $\dot{E}\approx \Omega^4\mu^2/c^3$.

This picture was, however, never verified by a first-principle calculation and 
was questioned
% (Michel 2004; Gruzinov 2014).
\citep{michel_state_2004,gruzinov_aristotelian_2013}.
It was shown that charges lifted from the star by the rotation-induced 
electric field form the ``electrosphere''
--- a corotating dome+torus structure,
with a huge gap between them and no electric current 
% (Jackson 1976; Krause-Polstorf \& Michel 1985). 
\citep{jackson_1976, krause-polstorff_electrosphere_1985}.
Although the electrosphere
is prone to diocotron instability 
% (Petri et al. 2001; Spitkovsky \& Arons 2002), 
% \citep{petri_diocotron_2002,spitkovsky_simulations_2002},
\citep{petri_diocotron_2002,2002ASPC..271...81S},
it was unclear if it could relax to the GJ state.

In addition to lifted charges, $e^\pm$ pairs are created around pulsars
% (Sturrock 1971). 
\citep{sturrock_model_1971}.
This provides plasma capable of 
screening the electric field component parallel to the magnetic field, $E_\parallel$.
The negligible plasma inertia and $E_\parallel=0$ provide the ``force-free'' (FF)
conditions, which imply GJ corotation.
The global solution for FF magnetospheres was obtained 
using various numerical techniques
% (Contopoulos et al. 1999; Timokhin 2006; Spitkovsky 2006; 
% Kalapotharakos \& Contopoulos 2009; Parfrey et al. 2012).
\citep{contopoulos_axisymmetric_1999,timokhin_force-free_2006,spitkovsky_time-dependent_2006,kalapotharakos_three-dimensional_2009,parfrey_introducing_2012}.
Its characteristic feature is a thin current sheet supporting 
a discontinuity of $\bB$ and the Y-point near the light cylinder.
It was verified with particle-in-cell (PIC) simulations that
sprinkling pairs with a high rate everywhere around the neutron star
would drive the magnetosphere to the FF configuration \citep{philippov_ab_2014}.

A self-consistent model must, however, demonstrate how and where the plasma is 
created and to identify the regions of $E_\parallel\neq 0$ (called ``gaps'') 
where particles are accelerated to high energies. Besides testing the FF
approximation, the self-consistent model would show how the pulsar 
radiation is produced, how the plasma flows in the magnetosphere and gets ejected.
This problem was posed soon after the discovery of pulsars and proved to be difficult.
Three types of gaps were proposed: polar-cap gap 
% (Sturrock 1971; Ruderman \& Sutherland 1975)
\citep{sturrock_model_1971,ruderman_theory_1975}, slot gap 
% (Arons 1983; Muslimov \& Harding 2004)
\citep{arons_pair_1983,2004ApJ...606.1143M}, 
and outer gap 
% (Cheng \& Ruderman 1986).
\citep{cheng_energetic_1986}.

The only reliable way to solve the problem is a first-principle calculation of the 
self-consistent dynamics of the electromagnetic field and pair discharge in the 
magnetosphere. Below we present such a direct numerical experiment.
Our simulations are performed with a new 2.5D PIC code, developed from 
scratch and designed for neutron-star magnetospheres.
The code calculates the fully relativistic dynamics of particles and fields
on a curvilinear grid, traces the emission of gamma-rays and their conversion to pairs.
The fields obey Maxwell equations, $\partial\bB/\partial t=-c\nabla\times\bE$ and 
$\partial\bE/\partial t=c\nabla\times\bB - 4\pi\bJ$, and exert force on particles 
$e(\bE+\bv\times\bB/c)$. We use 
% Esirkepov (2001)
\citet{esirkepov_2001}
charge-conserving scheme for calculating $\bJ$
and a semi-implicit algorithm for the field evolution.
A detailed description of the code and tests are given in the accompanying paper
(Chen \& Beloborodov, in preparation).

Pair creation by accelerated particles occurs in two steps: production of gamma-rays 
and their conversion to $e^\pm$. In many pulsars, the conversion is only efficient 
at $r\ll\RLC$ where the magnetic field is strong. In young fast pulsars pairs can 
be created through photon-photon collisions inside and around
the light cylinder 
% (Cheng \& Ruderman 1986). 
\citep{cheng_energetic_1986}.
For brevity, we call such rotators 
``type I.'' Pulsars with pair creation confined to $r\ll\RLC$ will be called type II.

%#############################################################
%#############################################################
\section{Problem formulation and simulation setup}

The axisymmetric pulsar is described by its radius $\Rs\approx 10$~km,
angular velocity $\bOm$, and magnetic dipole moment $\bmu$ (aligned or
anti-aligned with $\bOm$).  These parameters set the energy scale of
the problem. The neutron star is a nearly ideal conductor, and its
rotation induces voltage $\Phipc\approx \mu\Omega^2/c^2$ across the
footprint of the open field line bundle;
it corresponds to possible particle acceleration up to Lorentz factors
$\gamLC=e\Phipc/m_ec^2$. 
We start our simulations with $\Omega=0$ and the vacuum dipole field.
Then we gradually spin up the star: $\Omega$ grows linearly until 
it reaches its final value at $t_0=10 \Rs/c$;
$\Omega=const$ afterwards.

Corotational charge density 
$\rho_{\rm GJ}\approx -\boldsymbol{\Omega}\cdot \bB/2\pi c$
defines the characteristic particle density $n=|\rho_{\rm GJ}|/e$, plasma
frequency $\omega_p=(4\pi ne^2/m_e)^{1/2}$, and skin-depth
$\lambda_p=c/\omega_p$.  The magnetic field also determines the
gyro-frequency of $e^\pm$, $\omega_B=eB/m_ec$, and ions,
$\omega_{B,i}=eB/m_ic$. In a dipole magnetic field 
the characteristic frequencies 
are related by $\omega_p^2=2\omega_B\Omega$ and satisfy
$\Omega\ll\omega_p\ll\omega_B$.
The particle Larmor radius satisfies 
$\rL\ll r$ at $r\ll\RLC$, so particles move nearly along $\bB$. At the light cylinder,
$\rL\sim(\gamma/\gamma_0)\RLC$ may become comparable to $\RLC$.

The characteristic $\lambda_p$ at the polar cap is related to particle
acceleration, as $\gamLC\approx (1/4)(\Rs/\RLC)(\Rs/\lambda_p)^2$.
Typical pulsars have $\Rs/\lambda_p\sim 10^6\gg 1$ and $\RLC/\Rs\sim
10^2-10^3$.  We scale down the big numbers, preserving the hierarchy
of scales. The scale $\lambda_p$ must be well resolved in the
simulation, and the number of particles per grid cell must be large.
This can only be achieved by increasing $\lambda_p/\Rs$.
The simulations presented below have 
$\Rs/\lambda_p\approx 100-130$, $\RLC/\Rs=6-10$, and  $\gamLC=425$.
We also reduced the ion mass, $m_i=5m_e$,
and assumed ion charge number $Z=1$.

We use spherical coordinates $r,\theta,\phi$, and our grid is uniformly spaced 
in $\log r$ and $\theta$ to allow better resolution near the star, where it is most 
needed. The grid size is $512\times 512$.

Three field components are continuous at the star surface which defines
the boundary conditions: $B_r=2\mu\cos\theta/\Rs^3$, $E_{\theta} =
-\mu\Omega\sin 2\theta/\Rs^2$, and $E_\phi=0$.
The dipole configuration of the magnetosphere is set by the surface $B_r$, and 
its rotation is communicated from the star through the surface $E_\theta$.  
The star is also a source of electrons and ions for the magnetosphere.
We make both particle species available below the surface and their extraction
is self-consistently controlled by the local electric field.  We neglect the work function, 
so that particles are easily lifted from the star. 

The outer boundary is set at $\Rout=30\Rs\gg\RLC$. 
Here we place a ``damping layer'' of thickness $\Delta R=\Rout/15$.
The layer has resistivity and damps electromagnetic fields on
a timescale $\sim\Delta R/c$.  Particles escape freely; they
are decoupled from the fields once they enter the damping layer.  With
this implementation, the boundary effectively absorbs waves and
particles, which is equivalent to their free escape.

Once an electron or positron reaches the threshold energy $\gthr
m_ec^2$ it begins to emit curvature photons capable of pair creation.
It depends on the curvature radius of the particle trajectory $R_c$ as
$\gthr= K (R_c/\Rs)^{1/3}$. In real pulsars $K\simgt 10^6$; we scale
it down to $K=20$ to allow copious pair creation in our numerical
experiment.  
The photon emission rate is $\dot{N}=0.25 c(\gamma/R_c)$, 
where $\gamma$ is the particle Lorentz factor.
The photon emission, propagation, and conversion are
traced using Monte-Carlo technique.  The free paths of photons $l$
have a distribution $P(l)$ with mean value
$\lav$ and dispersion $\ldisp$. The extreme case of $\lav=0$ is only relevant for 
discharge near magnetars \citep{beloborodov_corona_2007};
for ordinary pulsars, the delay $l/c$ should be included. In our simulations
$\lav=\ldisp=2\Rs$ for photon-photon collisions
\AB{(operating in rotators of type~I)} 
and $\lav=\ldisp=0.2\Rs$ for magnetic conversion (enabled at $r\simlt 3\Rs$).
The emitted photons have energies $E_{\rm ph}\ll \gthr m_ec^2$, and hence
the secondary pairs are created with Lorentz factors $\gamma_s\ll\gthr$.
The condition $\gamma_s\ll \gthr\ll\gamLC$ ensures sufficient pair supply in the 
magnetosphere, and is satisfied in our simulations.
Radiation reaction (energy loss due to gamma-ray emission) is 
explicitly included in the particle dynamics.

Hereafter distance is measured in $\Rs$, time in $\Rs/c$, energy in $m_ec^2$,
magnetic and electric fields in $m_ec^2/e\Rs$, and
charge density in $m_e c^2/4\pi e\Rs^2$.

%###############################################################
%%%%%%%%%%%%%%%%%%%%%%%%%%%%%%%%%%%%%
\begin{figure}[t]
 %   \centering
 \hspace*{-0.6cm}
    \includegraphics[width=1.07\textwidth]{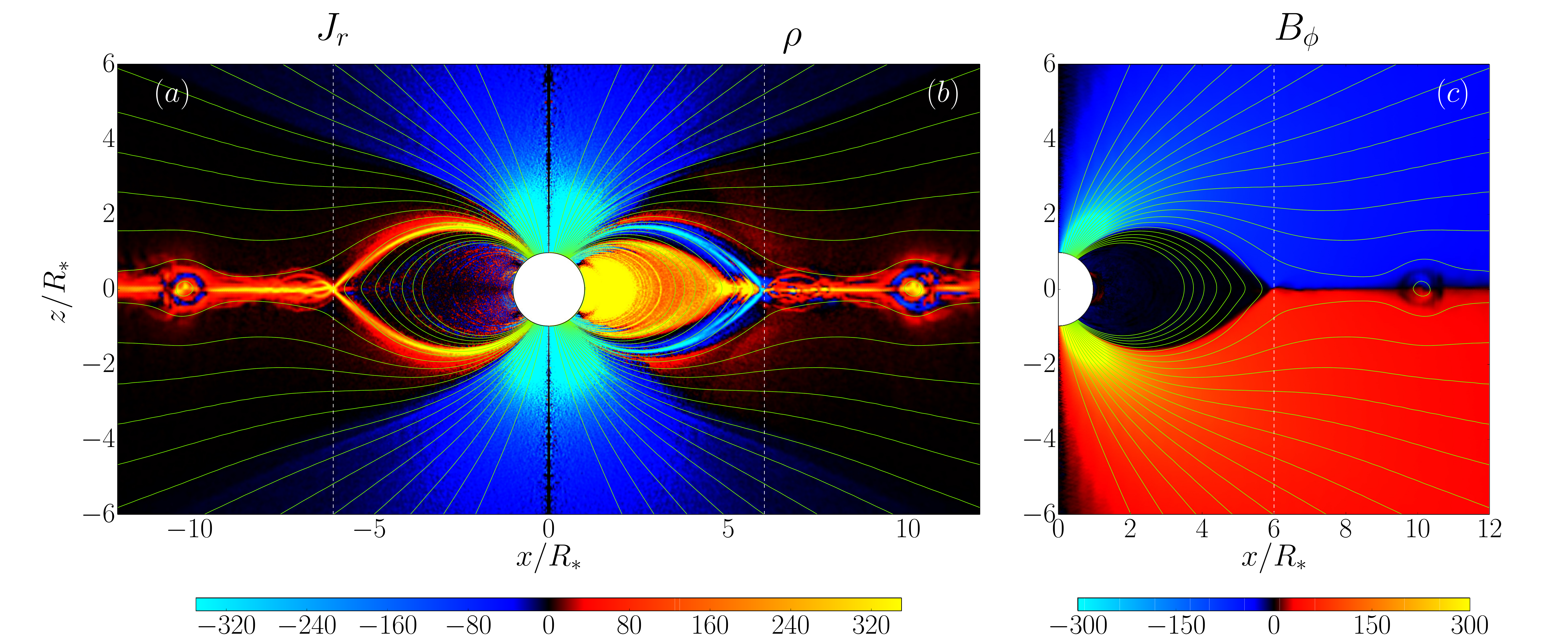}
    \caption{\small 
    Magnetosphere of type I aligned rotator (poloidal cross section) at $t = 100$.
    Vertical dashed line shows the light cylinder. 
    Green curves show the magnetic flux surfaces.
    \AB{(a) Radial component of electric current density $J_r$. 
           (b) Net charge density $\rho$.
           (c) Toroidal component of the magnetic field $B_\phi$.}
 }
    \label{fig:j-rho-bphi}
\end{figure}
%%%%%%%%%%%%%%%%%%%%%%%%%%%%%%%%%%%%%

\section{Rotators of type I}

\begin{figure}[t]
%    \centering
\hspace*{-0.6cm}
    \includegraphics[width=1.07\textwidth]{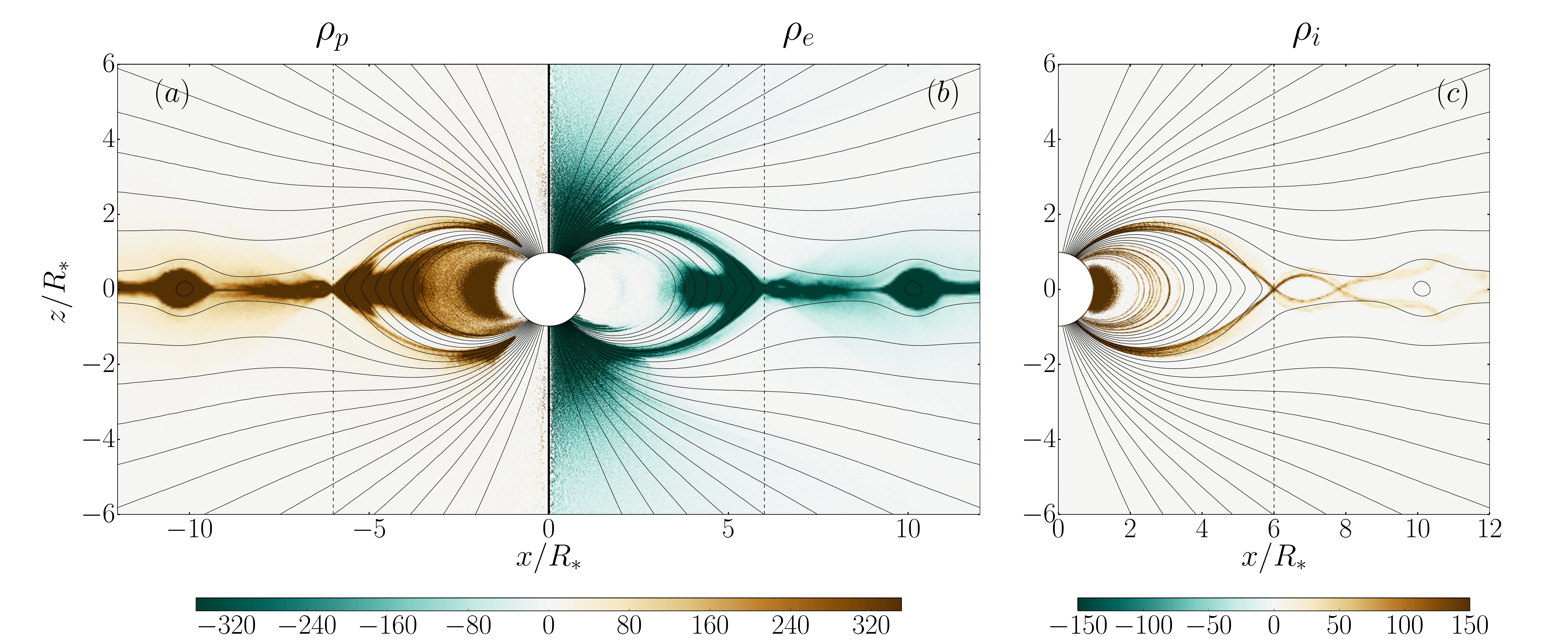}
    \caption{\small 
Charge densities of (a) positrons, (b) electrons, and (c) ions.
}
    \label{fig:rho-pn-corotation}
\end{figure}

\begin{figure}[t]
    \centering
    \includegraphics[width=0.85\textwidth]{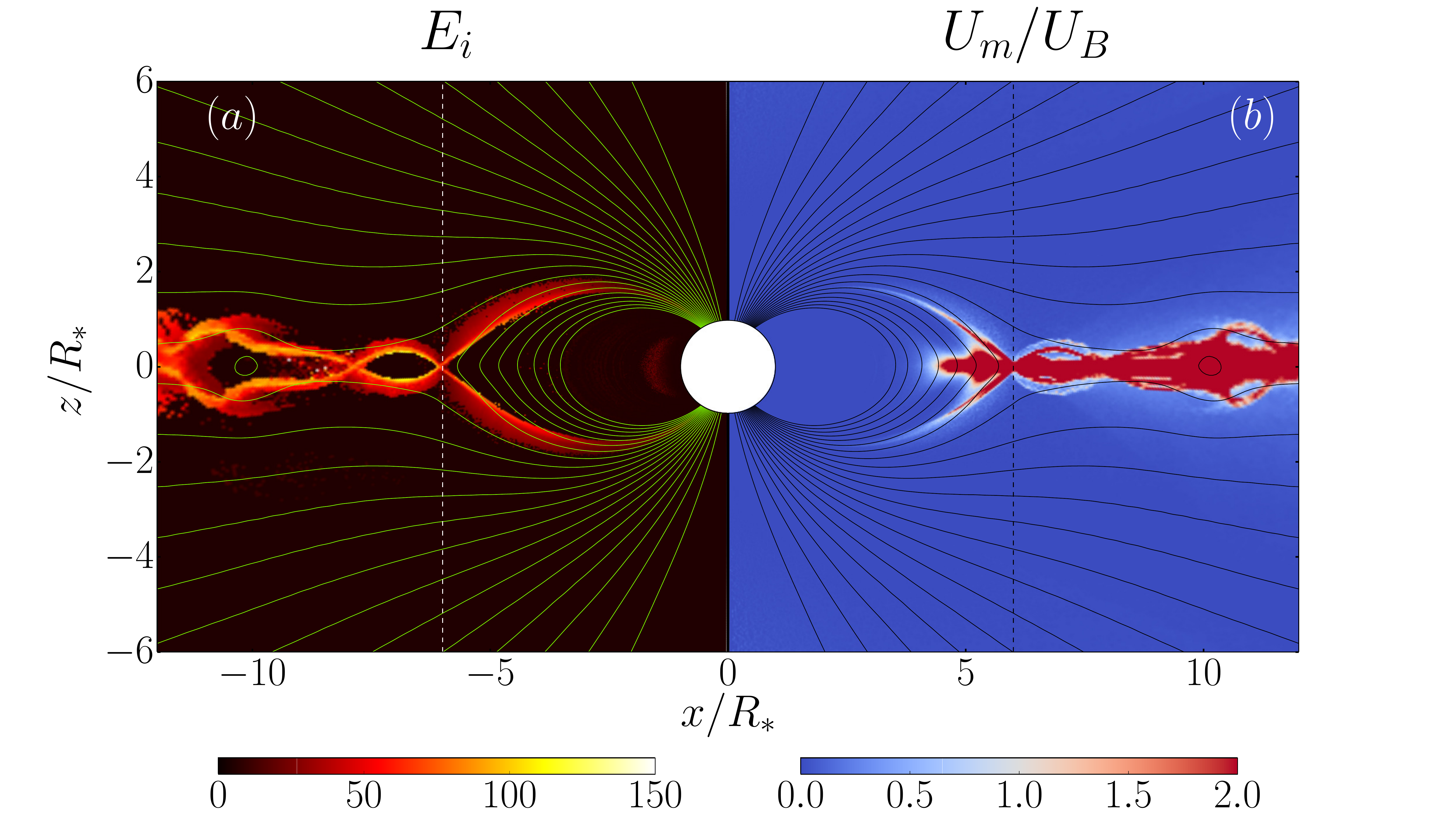}
    \caption{\small 
    (a) Average ion energy in units of $m_ec^2$. (b) Ratio of total matter energy density 
    $U_m$ to magnetic energy density $U_B=B^2/8\pi$.
}
    \label{fig:gamma-U-photon}
\end{figure}

Figures~\ref{fig:j-rho-bphi}--\ref{fig:gamma-U-photon} show the magnetosphere of the 
aligned rotator with $\RLC=6$ and $\mu=1.5\times 10^4$ after 2.6 rotation periods.
The energy density is almost everywhere dominated by the electromagnetic 
field, and the discharge finds a way to adjust and supply the charge density and 
electric currents demanded by the field. As a result, the magnetosphere 
shares several key features with the FF solution.
Electric currents and Poynting flux flow through the light cylinder along 
the open magnetic field lines while the interior of the closed field-line zone
has $\bJ=0$ and $B_\phi=0$. The Y-point is observed near $\RLC$.

There are two distinct regions of negative and positive radial current density $J_r$.
The negative current flows in the polar region around the magnetic axis.
The positive current is concentrated in a current sheet supporting
the jump of $B_\phi$ between the closed and open zones.
Outside the light cylinder, the current sheet extends along the equatorial plane 
to support the flip of $B_\phi$ and $B_r$ across the equatorial plane.

%%%%%%%%%%%%%%%%%%%%%%%%%%%%%%%%%%%%%
%%%%%%%%%%%%%%%%%%%%%%%%%%%%%%%%%%%%%

Charge density $\rho=\nabla\cdot \bE/4\pi$ 
also conforms to the expectations from the FF model
\citep[cf. Figure~16 in][]{parfrey_introducing_2012}.
In particular, the current sheet is positively charged outside the Y-point and 
negatively charged inside the Y-point 
(see Timokhin 2006 for discussion).
% \citep[see][for discussion]{timokhin_force-free_2006}.
$\rho$ significantly deviates from the FF model in the neutral black region with 
$J_r=0$; if the rotator approaches the ``death line''  for pair creation, 
$\gthr\sim\gamLC$, this region grows and occupies most of the magnetosphere.
A similar neutral
 region was described by 
% Yuki \& Shibata (2012).
 \citet{yuki_particle_2012}.

%%%%%%%%%%%%%%%%%%%%%%%%%%%%%%%%%%%%%
%%%%%%%%%%%%%%%%%%%%%%%%%%%%%%%%%%%%%

The two opposite currents are sustained by different mechanisms. 
The negative current in the polar region is carried by electrons lifted
from the polar cap. There is no significant activity in this region;
the particle acceleration is weak and pair creation does not occur. The absence
of polar-cap activity is explained by the low positive value of
$\alpha\equiv J_\parallel/c\rhoGJ\sim 0.7 <1$.
It leads to easy screening of $E_\parallel$ by the charge-separated
flow extracted from the star and the flow Lorentz factor
comparable to $2\alpha/(1-\alpha^2)$ 
% (Beloborodov 2008; Chen \& Beloborodov 2013).
\citep{beloborodov_polar-cap_2008,chen_dead_2013}.
We observed the same behavior in the simulation of anti-aligned rotator where currents
switch sign and the polar current is carried by ions extracted from the star.

The opposite current (the current sheet) is sustained by $e^\pm$ discharge
at $r<\RLC$.
It cannot be conducted by particles lifted
from the star as its sign is opposite to that of the
charge density demanded by the magnetosphere. Note also that 
$|\rho|\gg|\rhoGJ|$ in the current sheet, so $\rhoGJ$ 
is not important. The accelerating potential drop 
is $\Phi_\parallel\sim 2\pi \rho\delta^2 \sim -(\delta/r)\Phi_0$ where $\delta$ is the 
sheet thickness and we used $2\pi  r\delta |\rho| c\sim\IGJ=c\Phipc$.
Pair creation is biased to the outer side of the sheet (a result of its curvature 
and the finite free path of photons), therefore the unscreened $\Phi_\parallel$ is largest 
on the inner side. The sheet thickness $\delta$ is set by the Larmor radius of particles
near the Y-point.

Plasma outflows along the equatorial plane outside $\RLC$
and the Y-point resembles a nozzle formed by the open magnetic fluxes of opposite 
polarity. Two plasma streams come to the Y-point along the boundary of the closed 
zone and exchange their opposite $\theta$-momenta.
Their collimation is achieved through gyration in the (predominantly toroidal) 
magnetic field, which communicates the $\theta$-momentum from one stream to the other.
As a result the streams flow out in the direction of their net momentum, which is radial
% (see also Shibata 1985).
\citep[see also][]{shibata_1985}.

About $10$\% of current is carried by the ions 
extracted from the star at the footpoints of the current sheet.
Ions experience no radiative losses and tap the full $\Phi_\parallel$ 
(Figure~\ref{fig:gamma-U-photon}a). They have the largest Larmor radius
$\rL$, so the ion streams show large oscillations around the equatorial plane 
(Figure~\ref{fig:rho-pn-corotation}c). The streams with smaller oscillations are formed by 
accelerated positrons with $\gamma$ limited by radiation reaction.
Secondary particles have
even smaller energies; they outflow almost exactly in the equatorial plane.
The streams with different $\rL$ contribute to the thickening of the equatorial 
current sheet as seen in Figures~\ref{fig:j-rho-bphi}-\ref{fig:gamma-U-photon}.

Since ions do not create pairs, the discharge in the current sheet 
relies on the accelerated $e^\pm$. This requires continual recycling
of created particles as seeds for new rounds of pair creation, which
leads to voltage oscillations.
The oscillations occur on the timescale $\sim\RLC/c=\Omega^{-1}$ and make the 
magnetosphere ``breath'' around $\RLC$.

There is a steep potential drop across the outer closed field lines toward the
Y-point. The strong $E_r$ helps eject particles into the equatorial current sheet.
In this region $E\approx B$ and the particle ejection across $\bB$ is assisted by 
the drop of $B$ near the $Y$-point on a scale $\sim\rL$.

\begin{figure}[t]
   % \centering
    \hspace*{-0.6cm}
    \includegraphics[width=1.07\textwidth]{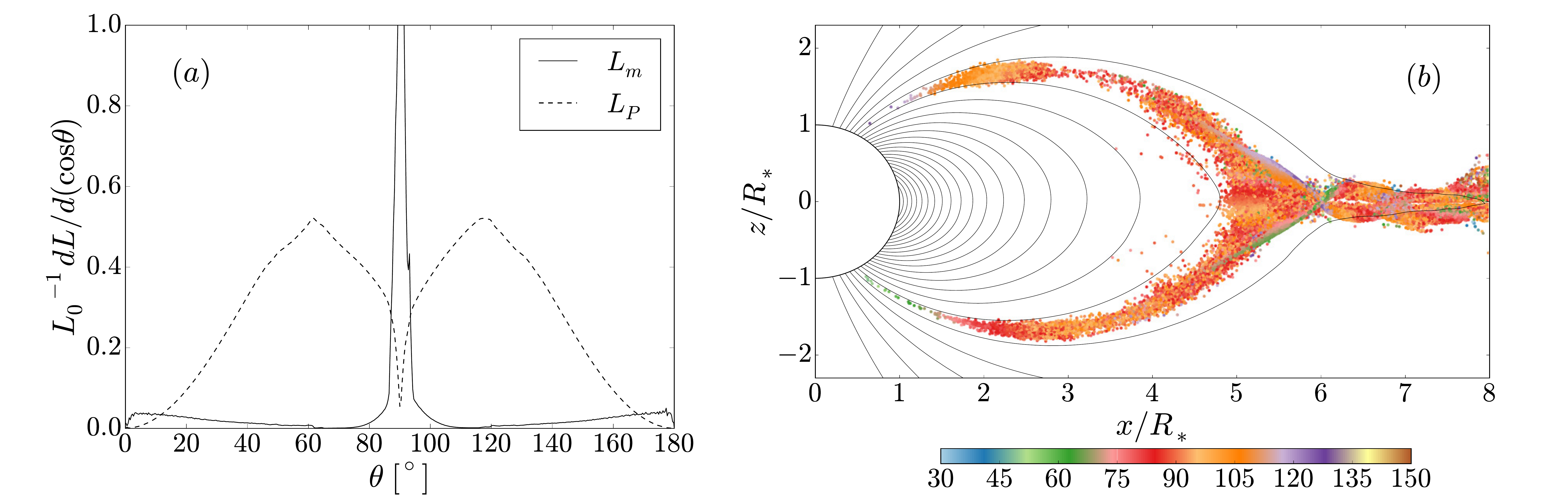}
    \caption{\small 
        (a) Angular distribution of
        Poynting flux $L_P$ and kinetic energy flux $L_m$ at radius
        $2\RLC$, normalized to  $L_0 = \mu^2\Omega^4/c^3$.
        (b) Locations of gamma-ray emission
        events. Color shows the angle of the photon 
        direction with respect to $\hat{\mathbf{z}}$. 
          }
    \label{fig:fluxes}
\end{figure}

The magnetic field dominates energy density everywhere except the Y-point region
and the matter-dominated equatorial outflow (Figure~\ref{fig:gamma-U-photon}b).
This behavior is also visible in the angular distributions of the Poynting luminosity 
$L_P$ and matter kinetic power $L_m$ (Figure~\ref{fig:fluxes}a).  
The integrated luminosities at $r=2\RLC$ are $L_P\approx 0.7 L_0$ and 
$L_m\approx 0.15 L_0$ where $L_0=\mu^2\Omega^4/c^3$.
Both contribute to the energy flux from the rotator.
This should be compared with the spindown power extracted from the star,
$\Lsd=L_P(\Rs)\approx 0.88L_0$.
The difference between $\Lsd$ and 
$L_P+L_m$ is carried by particles bombarding the star 
(the backflow power is $\sim 0.1L_m$) and the gamma-rays.

At $r\gg\RLC$ we observe magnetic reconnection which strongly heats 
particles in the equatorial outflow, forms large plasmoids, kinks and wiggles.
They are advected outward and do not affect the current structure near the Y-point.

The current sheet is the only source of gamma-rays (Figure~\ref{fig:fluxes}b).
The emission is strongly anisotropic, peaking at
$\pm \sim 12^{\circ}$ around the equatorial plane. In addition,
there is a strong peak at the equator from the high-energy
particles gyrating in the equatorial outflow.

%%%%%%%%%%%%%%%%%%%%%%%%%%%%%%%%%%%%
%%%%%%%%%%%%%%%%%%%%%%%%%%%%%%%%%%%%

The simulation for the anti-aligned rotator shows a similar magnetosphere
but somewhat different discharge.
The current sheet extracts 
and accelerates electrons from the star 
(instead of ions), which helps
produce pairs. This leads to a more stable $\Phi_\parallel$
and reduces the ``breathing'' of the magnetosphere, so the Y-point is nearly static. 

We also performed runs for the aligned and anti-aligned rotators with
different rotation rates, magnetic fields, and $\gthr$. The exact
position of the Y-point $R_Y$ depends on these parameters.  
With extremely low $\gthr$ and copious supply of plasma $R_Y$ decreases.
If $\gthr/\gamLC$ is increased to $\sim 1$, 
the magnetosphere transitions to the electrosphere state.

%###############################################################

\section{Rotators of type II}

The simulation of type~II rotators is the same as for type~I except for three changes: 
we suppressed pair creation where $B<400$, which roughly corresponds to $r\simgt 4$,
increased $\RLC$ to 10 to better separate it from the pair creation zone, 
and increased $\mu$ to $2.5\times 10^4$ to keep $\Phipc$ unchanged.

\begin{figure}[t]
    \centering
    \includegraphics[width=0.85\textwidth]{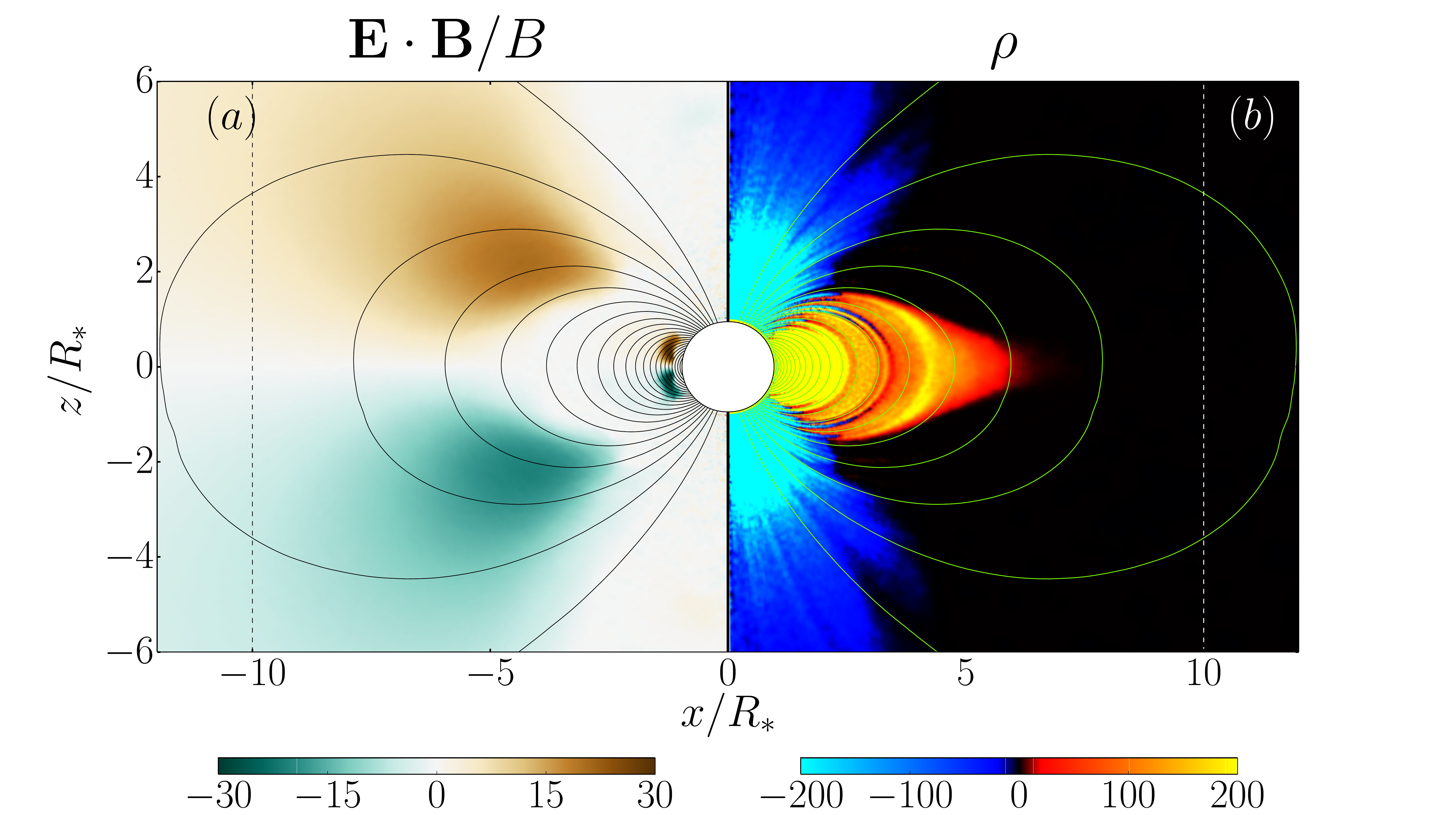}
    \caption{\small 
    Type II magnetosphere at $t=280$. 
    (a) Parallel electric field $E_{\parallel}$.
    (b) Charge density $\rho$.
    }
   \label{fig:type2}
\end{figure}

As we spin up the star, there is an initial burst of pair creation due to the
vacuum initial condition and the induced $E_\parallel$ accelerating
charges extracted from the surface. The system is able to form an almost FF
configuration for a short time, then $E_\parallel$ gets screened inside the 
pair-creation zone, and the magnetosphere relaxes to a different state.
After about 2 rotations of the star, $\rho\approx\rhoGJ$ is sustained only at $r\simlt 4$,
and outside this zone the magnetosphere is close to the electrosphere solution 
(Figure~\ref{fig:type2}). It stays in this state till the end of the simulation 
(about 8 rotations), with no pair creation.

%%%%%%%%%%%%%%%%%%%%%%%%%%%%%%
%%%%%%%%%%%%%%%%%%%%%%%%%%%%%%

In the final state, the magnetic field is everywhere similar to the dipole.
The plasma forms the negative ``dome'' and positive ``torus'' 
with a vacuum gap between them at $r\simgt 4$. 
The gap is outside the pair-creation zone and finds no way for plasma supply.
The unscreened $E_\parallel$ in the gap creates a 
large potential drop along the 
magnetic field lines, which leads to faster rotation of the magnetosphere in the 
equatorial region at $r\sim 5-8$ 
% \AB{(cf. Wada \& Shibata 2011).}
\citep[cf.][]{wada_shibata_2011}.

The magnetosphere is not completely dead at the end of the simulation.
There is a weak negative current flowing out in the polar cap region, and
a weak positive current leaking out from the tip of the ``torus'' region
due to a strong $E_r\approx B$.
At later times we expect the magnetosphere to evolve even closer to
the electrosphere solution.  The gap will tend to expand toward the
null point $\rhoGJ=0$
on the star surface, into the zone where pair creation is
possible. Then the discharge should reignite and prevent the inward
expansion of the gap. The duration of our simulation is too short to
see this late evolution; however, it is sufficiently long to show that type
II rotators have a low level of activity most of the time.

Breaking the axisymmetry should initiate the diocotron instability
% (e.g. Philippov \& Spitkovksy 2014); 
\citep{philippov_ab_2014};  however it is unlikely to 
prevent the inner zone from filling with plasma and 
shutting down pair creation. Our simulations suggest that type II rotators
cannot relax to the GJ state
--- it is unstable even with suppressed azimuthal perturbations; allowing the 
diocotron instability cannot make the GJ state an attractor for the system.

%################################################################

\section{Discussion}

Our first conclusion is that significant activity of the axisymmetric pulsar
requires pair creation enabled at $r\sim\RLC$ 
(called type I in this paper); this requires a sufficient optical 
depth to photon-photon collisions.
If pair creation is limited to $r\ll\RLC$ (type II), 
the return current is choked and the axisymmetric magnetosphere relaxes to the 
dome+torus state with suppressed electric currents and pair creation.  
Many observed pulsars are only capable of pair creation at
$r\ll\RLC$; we conclude that their activity and spindown should
be a result of the misalignment of $\bOm$ and $\bmu$. This conclusion
supports the arguments of
% Michel (2004)
\citet{michel_state_2004}
 and disagrees with the models of 
% Goldreich \& Julian (1969), 
\citet{goldreich_pulsar_1969},
% Ruderman \& Sutherland (1975),
\citet{ruderman_theory_1975},
and 
% Gruzinov (2014).
\citet{gruzinov_aristotelian_2013}.

Type I axisymmetric rotators are active, as long as discharge voltage 
$\Phi_{\rm thr}<\Phi_0=\mu\Omega^2/c^2$.
Their spindown power is $\Lsd\approx \mu^2\Omega^4/c^3$ 
and their magnetic configuration is similar to the FF solution.
Our numerical experiment shows, for the first time, 
how particle acceleration and $e^\pm$ discharge 
self-organize to maintain this configuration. 
The result is quite different from the 
previously discussed ``trio'' of gaps: polar-cap gap, slot gap, and outer gap.
Neither aligned nor anti-aligned rotators sustain pair creation 
in the polar cap outflow.
Strong particle acceleration and pair creation occur in (and around) 
the return current sheet
stretched along the boundary of the closed zone.  
The acceleration mechanism is different from
the slot-gap models, which were developed for the opposite,
polar-cap current. It is also different from the outer gap model where
the null surface $\rhoGJ=0$ plays a key role.  
We find that the current sheet has
$|\rho|\gg|\rhoGJ|$, and its $E_\parallel$ is not controlled by the
local value of $\rhoGJ$.

Our numerical experiment confirms the phenomenological description of the 
gamma-ray source as an accelerator stretched along the boundary of the closed 
zone, which explains the observed pulse profiles of GeV emission 
% (Dyks \& Rydak 2003). 
\citep{dyks_two-pole_2003}.
The angular distribution of gamma-ray luminosity is determined by
how $E_\parallel$ and $e^\pm$ discharge self-organize to sustain the magnetospheric
configuration; the geometry by itself does not determine this distribution.

Besides producing copious $e^\pm$ pairs, type I rotators eject a significant flux 
of ions, $\dNi$. The anti-aligned rotator ejects $\dNi\approx \IGJ/e$ 
from the polar cap, with low energies. The aligned rotator in our simulation ejects
$\dNi\sim 0.1\IGJ/e$ along the current sheet, with much higher energies, which 
carries $\sim 5$\% of the spindown power.

It remains to be seen which features of the axisymmetric magnetosphere
will hold for inclined rotators.
FF models provide guidance as they show where the current should 
flow. In contrast to the axisymmetric case, inclined rotators have $\alpha<0$ and 
$\alpha>1$ in the central region of the polar cap, which is required to activate 
$e^\pm$ discharge
% (Beloborodov 2008; Chen \& Beloborodov 2013; Timokhin \& Arons 2013).
\citep{beloborodov_polar-cap_2008,chen_dead_2013,timokhin_current_2013}.
A current sheet with $|\alpha|\gg 1$ is expected to form along the boundary of 
the closed zone and produce gamma-rays similar to the mechanism seen in our
simulations.

Our results show that key puzzles in pulsar physics
can be solved using first-principle calculations, opening exciting opportunities
for future modeling. This includes the global magnetic configuration, particle 
acceleration,
pair multiplicity, and broad-band radiation, from 
curvature gamma-rays to coherent radio waves.

\acknowledgements 
This work was supported by NASA grant NNX~13AI34G. 
We thank Beno\^{i}t Cerutti, Andrei Gruzinov, Kyle Parfrey, and Anatoly Spitkovsky
for helpful discussions.

%##############################################################

\end{document}